\documentstyle[aps,epsfig,twocolumn]{revtex}

\begin{document}

\twocolumn[\hsize\textwidth\columnwidth\hsize\csname@twocolumnfalse\endcsname

\title{Absence of the zero bias peak in vortex 
tunneling spectra of high temperature superconductors}

\author{Congjun Wu$^{1,2}$, Tao Xiang$^{2}$ and Zhao-Bin 
Su$^{2}$}

\address{$^{1,2}$Department of Physics, Peking 
University, Beijing 100087, The People's Republic of China}

\address{$^2$Institute of Theoretical Physics, 
Chinese Academy of Sciences, P.O.Box 2735, Beijing 100080, 
The People's Republic of China}

\maketitle

\begin{abstract}
The $c$-axis tunneling matrix of high-$T_c$ superconductors 
is shown to depend strongly on the in-plane momentum of electrons and 
vanish along the four nodal lines of the $d_{x^2-y^2}$-wave energy gap. 
This anisotropic tunneling matrix suppresses completely the 
contribution of the most extended quasiparticles in the vortex 
core to the $c$-axis tunneling current and leads to a 
spectrum similar to that of a nodeless superconductor. 
Our results give a natural explanation of the absence of the zero 
bias peak as well as other features observed in the vortex tunneling 
spectra of high-$T_c$ cuprates.
\end{abstract}

]

Recently there has been intensified attention on the 
electronic structure in the $d_{x^2-y^2}$-wave vortex 
state of high-T$_c$ superconductors 
\cite{magg,renn,pan,soin,wang,Morita,aro,franz1,yasui,and,Ichioka,franz2,lee}. 
From the solution of the 
Bogoliubov-de Gennes (BdG) equation, many authors 
\cite{wang,franz1,yasui} showed that in a $d$-wave 
superconductor there are no vortex core bound states and 
the local density of states (LDOS) at the core center has 
a broad zero energy peak. However, in contrast to the 
theoretical prediction, the scanning tunneling spectrum 
(STS), which is generally assumed to be proportional to 
the LDOS, at the center of a vortex core in both 
YBa$_2$Cu$_3$O$_{7-\delta }$ (YBCO)\cite{magg} and 
Bi$_2$Sr$_2$CaCu$_2$O$_{8+\delta }$ (BSCCO) 
\cite{renn,pan} have revealed a number of features which 
are totally unexpected in a pure d-wave superconductor: 
first, there is zero peak at zero bias; second, the 
coherent peaks at the superconducting gap edges are 
largely suppressed; and third, as in a $s$-wave 
superconductor, localized vortex core states seem to 
exist at $\pm5.5meV$ in YBCO \cite{magg} and $\pm 7meV$ 
in BSCCO \cite{pan}.

Many theoretical concepts, including spin-charge 
separation, SO(5) symmetry, and time reversal symmetry 
breaking pairing states have been invoked to resolve this 
discrepancy between theories and experiments 
\cite{aro,and,franz2,lee}. A simple theoretical 
explanation for the vanishing zero bias peak is that the 
pairing state inside the core has $d_{x^2-y^2}+id_{xy}$ 
or $d_{x^2-y^2}+is$ wave symmetry and therefore as in a 
$s$-wave superconductor vortex bound states exist in 
these materials \cite{franz1}. However, there is no direct 
experimental evidence for the existence of such 
time-reversal symmetry broken states. In the SO(5) 
theory the vortex core can be either metallic or insulating 
depending on the doping level or other physical parameters 
of a high-Tc material\cite{aro,and}. This is consistent with the STS data of 
YBCO \cite{magg} and BSCCO \cite{renn,pan}. Under the 
framework of the U(1) gauge theory, Franz et al 
\cite{franz2} argued that the pseudogap behavior of the 
STS at the core center is a result of the 
charge-spin separation and what the experiments found is 
the excitation spectrum of paired spinons which are not 
affected by external fields.

Nearly all theoretical analyses of the STS at the 
vortex core have assumed that the tunneling matrix 
element is a momentum independent constant. 
Under this assumption it can be shown that the 
c-axis tunneling differential conductance measured in the 
STS experiments is proportional to the LDOS of low lying 
excitations. However, for high-$T_c$ cuprates, the c-axis 
hopping integral of electrons is highly anisotropic 
\cite{Andersen,xiang1,char} and depends strongly 
on the in-plane momentum. In particular, it has the same 
nodal structure as the $d_{x^2-y^2} $-wave gap function 
in tetragonal cuprates like BSCCO. The tunneling 
matrix is proportional to the c-axis velocity of 
electrons, thus it should also depend strongly on 
the in-plane momentum of electrons. This anisotropic 
c-axis hopping integral has strong impact on the c-axis 
transport of quasiparticles in the superconducting state. 
It leads to some peculiar temperature dependence of the 
$c$-axis penetration depth\cite{xiang1,Pana} as 
well as microwave conductivity\cite{xiang3} at low 
temperatures and suppresses significantly the c-axis 
tunneling conductance \cite{franz3}.

In this paper, we present a theoretical analysis of the 
STS in the vortex core of high-$T_c$ superconductors. We 
shall show that the c-axis STS is dramatically modified 
by the anisotropy of the tunneling matrix. Our work 
provides a natural explanation for the absence of the 
zero bias peak in the STS at the core center. It shows 
that the pseudogap feature of the tunneling conductance 
at the core center is mainly due to the suppression of 
the tunneling current of low energy quasiparticles by the 
anisotropic c-axis hopping integral. A critical test for 
our theory is to measure the tunneling conductance in a 
vortex core which is parallel to the CuO$_2$ planes. The 
tunneling matrix parallel to the CuO$_2$ planes is not 
sensitive to the pairing symmetry and thus can be taken 
approximately as a constant. In this case, a broad zero 
bias peak is expected to exist in the STS of the vortex. 
This anisotropic feature of the STS is absent in other 
theories.

To study the structure of a $d$-wave vortex along the 
$c$-axis, we have performed a self-consistent calculation 
of the Bogoliubov-de Gennes (BdG) equation on a square 
lattice. Before presenting the detailed numerical 
results, let us briefly discuss why there is a broad 
zero-energy peak in the LDOS of a $d_{x^2-y^2}$-wave 
vortex and how it is suppressed in the scanning tunneling 
experiment by the anisotropic $c$-axis hopping integral.

For conventional s-wave superconductors, it was 
established many decades ago\cite{Caroli} that discrete 
quasiparticle states with a characteristic excitation 
energy given by $\Delta ^2/2\varepsilon _F$ exist in a 
vortex core, where $\Delta $ is the bulk gap and 
$\varepsilon _F$ is the Fermi energy. Intuitively, this 
can be understood by drawing an analogy to a simple 
quantum mechanical problem of a particle in a
cylindrical well of radius $\xi \sim v_F/\pi \Delta $ 
and height $\Delta $. However, in a $d_{x^2-y^2}$-wave 
superconductor, the radius and height of the analogous 
potential well depends on the polar angle on the Fermi 
surface. Along the four node directions, $\Delta $ 
vanishes and $\xi $ diverges. In this case the 
quasiparticle is extended along the node directions and 
there is no truly localized core states. Thus unlike the 
s-wave case, the DOS of quasiparticles in the vortex of 
$d_{x^2-y^2}$-wave superconductors is finite even at zero 
energy. From the numerical solution of the BdG equation 
for a $d_{x^2-y^2}$-wave superconductor, it has been 
further shown that there is a broad peak at zero energy 
in the LDOS of the vortex core. This peak mainly arises from 
the quasiparticle excitations along the four 
node lines.

The STS is determined by the LDOS and the tunneling matrix elements. 
In high-$T_c$ materials, since the $c$-axis tunneling 
matrix vanishes along the four node lines, the 
contribution of the most extended quasiparticles, around 
the gap nodes, to the tunneling current is  
completely suppressed. This is why the zero 
energy peak in the LDOS is absent in the STS of the vortex.

The BdG equation for a $d_{x^2-y^2}$-wave superconductor 
on a lattice is given by\cite{Gennes} 
\begin{equation}
\left( 
\begin{array}{ll}
\widehat{H}_0 & -\widehat{\Delta } \\ 
-\widehat{\Delta }^{*} & -\widehat{H}_0^{*}
\end{array}
\right) \left( 
\begin{array}{l}
u_n(i) \\ 
v_n(i)
\end{array}
\right) =E_n\left( 
\begin{array}{l}
u_n(i) \\ 
v_n(i)
\end{array}
\right)
\end{equation}
and 
\begin{eqnarray}
\widehat{H}_0u_n(i) &=&-t\sum_\delta u_n(i+\delta )-\mu 
u_n(i),  \nonumber \\
\widehat{\Delta }v_n(i) &=&\sum_\delta \Delta _{i,\delta 
}v_n(i+\delta ),
\end{eqnarray}
where $\delta $ denotes a nearest-neighbor vector, $t$ is 
the hopping constant, and $\mu $ is the chemical 
potential. $\left( u_n(i),v_n(i)\right)$ is the wave 
function of Bogoliubov quasiparticles. We have ignored 
the coupling to the vector potential $A$ in the hopping 
term in the limit of an extremely type II superconductor. 
The effect of the gauge field is manifested in the 
winding number of the order parameter in the calculation. 
The self-consistent condition for the gap parameter is 
\begin{equation}
\Delta _{i\delta }={\frac g2}\sum_n\left[ 
u_n(i)v_n^{*}(i+\delta
)+u_n(i+\delta )v_n^{*}(i)\right] \tanh {\frac{{\beta 
E_n}}2},
\label{charge}
\end{equation}
where $g$ is the pair coupling constant.

\begin{figure} 
\centering\epsfig{file=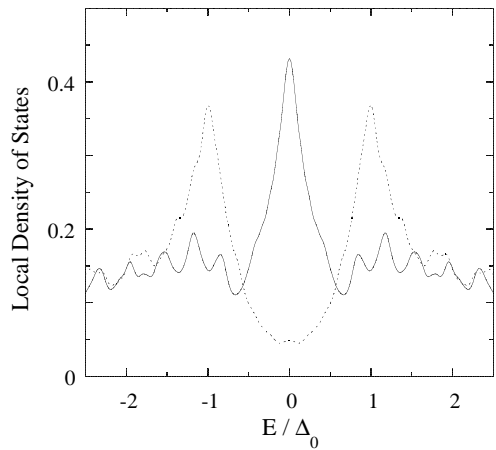,clip=,width=7cm,angle=0}
\caption{
Local density of states at the center of the vortex at 
site (0, 0) (solid line) and at site (10, 10) (dotted 
line) on a $51\times 51$. The small oscillations of the curves 
are due to the finite size effects. The energy is normalized 
by the maximum energy gap $\Delta_0=0.6t$. The broadening 
parameter is $\gamma = 0.05t$.
}
\end{figure}

The LDOS is determined by the wavefunctions of 
quasiparticles:
\begin{eqnarray}
\rho (i,E) &=&\sum_n\rho (i,n,E) \\
\rho (i,n,E) &=&|u_n(i)|^2L(E-E_n)+|v_n(i)|^2L(E+E_n),
\end{eqnarray}
where $L(x)=\gamma /\pi \left( {x^2+\gamma ^2}\right) $ 
and $\gamma$ an energy broadening parameter. $\rho 
(i,E)$ is a sum of the LDOS of the $n^\prime$th vortex 
eigenstate at site $i$, $\rho (i,n,E)$. Figure 1 shows 
the LDOS at site $\left( 0,0\right) $ (core center) and 
site $\left( 10,10\right) $ on a $51\times 51$ square 
lattice with open boundary conditions. The parameters 
used are $g=1.5$ and $\mu =0$. In agreement with other 
numerical results\cite{wang,franz1}, we find that the 
LDOS at the core center exhibits a broad peak at $\omega 
\sim 0$. This peak drops very quickly with the distance 
from the core center. The LDOS at site $\left( 
10,10\right) $ behaves in nearly the same was as in an 
homogeneous $d_{x^2-y^2}$-wave superconductor. There is a 
coherent peak at the gap edge ($\left|\omega 
\right| \sim \Delta _0$) at $\left( 10,10\right) $. 
However, at the core center, this coherent peak is 
suppressed.

The tunneling current between a normal tip and a 
superconductor at a bias voltage $V$ is given by
\begin{eqnarray}
I(i,V) &=&2e\sum_{n,p}|T_{nk_z,p}|^2\int_{-\infty 
}^{+\infty }{\frac{{%
d\varepsilon }}{{2\pi }}}A_S(n,i,\varepsilon )  \nonumber 
\\
&&A_N(p,\varepsilon +eV)\left[ n_F(\varepsilon )-
n_F(\varepsilon +eV)\right]
\label{Current}
\end{eqnarray}
where $A_S(n,i,\varepsilon )=2\pi \rho (i,n,\varepsilon 
)$ and $%
A_N(p,\varepsilon )=2\pi \delta (\varepsilon -\xi _p)$ 
are the spectral functions of the superconductor and 
normal tip, respectively. $p$ is the momentum of 
electrons in the normal tip and $(n,k_z)$ are the quantum 
numbers of the vortex states. The tunneling matrix element
$T_{nk_z,p}$ is proportional to the velocities of 
electrons along the tunneling direction on both sides of 
the junction \cite{wolf}: 
\begin{equation}
|T_{k_zn;p}|^2=D(\varepsilon _z)\left| 
v_p^nv{_{k_z,n}^s}\right| ,
\label{TunMax}
\end{equation}
where $D(\varepsilon _z)$ is the barrier transmission 
coefficient. $v_p^n$ is the Fermi velocity in the normal 
tip, which can be approximately taken as a constant. The 
{velocity of the }$n^{\prime }$th vortex state 
$v_{k_z,n}^s$ is determined by the energy dispersion of 
quasiparticles along the $c$-axis. To the lowest order 
approximation of $t\perp$ (defined below), 
it is proportional to the expectation 
value of the $c$-axis current operator in the $n^{\prime 
}$th vortex state.

For tetragonal high-$T_c$ compounds, the $c$-axis energy 
dispersion of electrons has been shown to have the form 
\begin{equation}
\varepsilon _z\left( k\right) =-t_{\perp }\left( \cos 
k_x-\cos k_y\right)
^2\cos k_z,  \label{chop}
\end{equation}
where $t_{_{\perp }}$ is the $c$-axis hopping constant. 
This anisotropic $c$-axis hopping integral results from 
the hybridization between antibonding $O$ $2p$ and 
unoccupied $Cu$ $4s$ orbitals which act as intermediate 
states\cite{xiang1}. It is also a good 
approximation for YBa$_2$Cu$_3$O$_{7-x}$\cite{Andersen}.

Eq. (\ref{chop}) can be obtained from the following 
interlayer hopping Hamiltonian 
\[
H_c=-\frac{t_{\perp }}8\sum_{m,\delta ,\delta ^{\prime 
}}D_\delta D_{\delta
^{\prime }}c_{i+\delta ,m}^{\dagger }c_{i+\delta ^{\prime 
},m+1}+h.c. 
\]
where $m$ is the index of CuO$_2$ planes and $D_\delta 
=1$ or $-1$ if $\delta =\pm \hat{x}$ or $\pm \hat{y}$. In 
a vortex system, since the translational symmetry is 
broken and $\left( k_x,k_y\right) $ are no longer good 
quantum numbers, it is more convenient to use this real 
space form of the c-axis hopping integral. The $c$-axis 
current operator is now defined by 
\[
J_c=\frac{it_{\perp }}8\sum_{m,\delta ,\delta ^{\prime 
}}\left( D_\delta
D_{\delta ^{\prime }}c_{i+\delta ,m}^{\dagger 
}c_{i+\delta ^{\prime
},m+1}-h.c.\right) . 
\]
From the expectation value of $J_c$ in the vortex state, 
we find $v_{k_z,n}^s$ to be 
\begin{equation}
\left| v_{k_z,n}^s\right| \sim t_{\bot }V_n\left| \sin 
k_z\right| ,
\label{ve}
\end{equation}
where 
\[
V_n=\sum_i\left[ \left|\sum _\delta D_\delta u_n(i+\delta 
)\right|^2+\left|\sum _\delta 
D_\delta v_n(i+\delta )\right|^2\right] 
\]
is the $c$-axis velocity factor of the $n^\prime$th vortex 
state.

Substituting Eqs. (\ref{TunMax}) and (\ref{ve}) into 
(\ref{Current}) and integrating out $k_z$, we find that 
the tunneling differential conductance is 
\begin{equation}
g(i,V) \sim \sum_nV_n\rho (i,n,eV).  \label{new}
\end{equation}
If $V_n$ does not depend on $n$, such as in an ordinary 
superconductor, $g(i,V)$ is simply proportional to the 
LDOS $\rho (i,eV)$. However, as discussed below, in high-
$T_c$ cuprates $V_n$ varies strongly with $n$. In this 
case $g(i,V)$ is completely different to $\rho (i,eV)$.

\begin{figure} \label{v}
\centering\epsfig{file=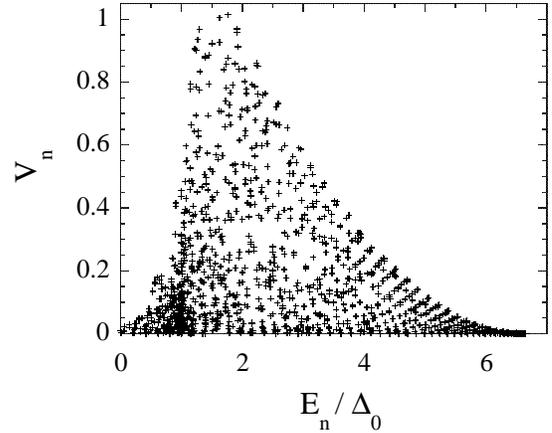,clip=,width=7.5cm,angle=0}
\caption{ The $c$-axis velocity factor $V_n$ as a 
function of energy. 
}
\end{figure}

Figure 2 shows the energy dependence of $V_n$ for the single 
vortex system discussed previously. The variance of 
$V_n$ is highly non-monotonic. At low energies, 
\thinspace $V_n$ is very small. It reaches a maximum when 
$E_n$ is slightly above the maximum energy gap $\Delta 
_0$. At high energies, $V_n$ drops with increasing 
energy. The behavior of $V_n$ in different energy regions 
can be understood from the behavior of the expectation 
value of $\left( \cos k_x-\cos k_y\right) ^2$ in each 
vortex eigenstate. At low energies, $V_n$ is small since 
it arises mainly from excitations around the 
gap nodes. In this energy region, $V_n$ has roughly the 
same energy dependence as $\left( \cos k_x-\cos 
k_y\right) ^2$ and its local maximum increases 
approximately with $E^2$. $V_n$ becomes large when $E_n$ 
is approximately equal to $\Delta _0$. This is because 
the maximum gap corresponds to the contribution of the  
quasiparticle around $(\pi ,0)$ 
where $\left( \cos k_x-\cos k_y\right) ^2$ 
is maximum. The high energy eigenstates are mainly 
from the excitations around $\left( 0,0\right) $ and 
$\left( \pm \pi ,\pm \pi\right)$. $V_n$ drops at high 
energies since $\left( \cos k_x-\cos
k_x\right) ^2=4\sin ^2\left( k_x+k_y\right) /2\sin 
^2\left( k_x-k_y\right)/2 $ and $\sin ^2\left( 
k_x+k_y\right) /2$ is zero at $\left( 0,0\right) $ and 
$\left( \pm \pi ,\pm \pi \right) $.

Figure 3 shows the tunneling differential conductance for 
the system studied above. The conductance at $\left( 
10,10\right) $ behaves in qualitatively the same way as for the 
LDOS, but its coherent peak at the gap edge is 
suppressed. At site $\left( 0,0\right) $ (the core 
center), the conductance differs completely from the 
LDOS. In particular, three distinct features appear in 
the tunneling conductance. (1) The $c$-axis velocity 
$V_n$ suppresses completely the zero energy peak of LDOS 
and leads to a small tunneling 
conductance at zero bias. The zero bias conductance is finite in the 
figure. This is due to the broadening of the spectrum by 
the parameter $\gamma $. In an infinite lattice system where 
$\gamma$ can be asymptotically set to zero, 
the value of the zero bias conductance should become zero since 
the tunneling matrix vanishes at the gap nodes. 
(2) There are two weak peaks at low bias. The positions of these 
weak peaks depend on $\Delta_0$. For the case shown in Figure 2, 
they are located at  $V\sim\pm 0.25\Delta_0$.
These peaks result from the competition between the decreasing 
density of states and the increasing velocity factor (on average) 
at low energies. Unlike the irregular oscillations caused 
by the finite size effects, the positions of these peaks do 
not change as the lattice size is increased. This shows that these 
weak peaks are not due to the finite size effects. The peak value is 
very small compared with the conductance at the gap edge 
$\left| V\right| \sim \Delta _0$, but it slightly 
increases with increasing lattice size. The existence of 
these low bias peaks with the absence of the zero bias 
peak is reminiscent of the tunneling spectrum of a 
conventional $s$-wave superconductor, although no vortex 
core bound states exist in this case. (3) There is no 
coherent peak at the gap edge. This is different than in 
a conventional $s$-wave superconductor. All these three 
features agree well with the experimental observations 
\cite{magg,renn,pan}.

\begin{figure}
\centering\epsfig{file=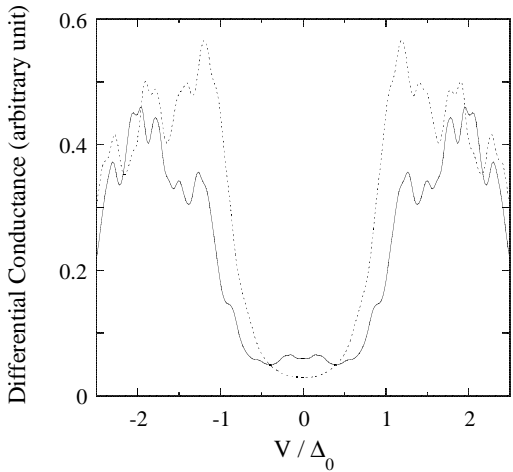,clip=,width=7cm,angle=0}
\caption{
Tunneling differential conductance at the core center at 
site (0,0) (solid line) and at site (10,10) (dotted line) 
for the same system as shown in Figure 1. 
}
\end{figure}

In the above discussion, particle-hole symmetry has 
been implicitly assumed since the chemical potential $\mu 
$ has been set to zero. For a particle-hole asymmetric 
system ($\mu \neq 0)$, we find that the differential 
conductance becomes asymmetric with respect to zero bias. 
However, the other features of the tunneling spectrum 
seen in the particle-hole symmetric system are 
unchanged. Therefore, we believe the asymmetry of the STS 
observed in the experiments is mainly due to the breaking 
of particle-hole symmetry \cite{magg,renn,pan}.

In conclusion, we have presented a theoretical 
analysis of the vortex tunneling spectra of high-$T_c$ 
superconductors. Our work indicates that it is important 
to include the anisotropy of the c-axis hopping integral 
in the analysis of the STS in high-$T_c$ oxides. The 
tunneling matrix elements parallel to CuO$_2$ planes do not have 
the same anisotropy as the $c$-axis case. Thus STS for 
a vortex perpendicular to the $c$-axis is expected to 
behave very differently. In particular, the zero energy 
peak of the LDOS at the core center should appear in the 
STS in this case. Thus by measuring the STS parallel to 
CuO$_2$ planes, we can obtain a better 
understanding of the $c$-axis tunneling experiments. 

We wish to thank S. Pan and C. Panagopoulos 
for useful discussions. This 
work was supported in part by the National Natural 
Science Foundation of China and by the Special funds for 
Major State Basic Research Projects of China.

\end{document}